\documentclass[pdftex,twocolumn,epjc3]{svjour3}   
\usepackage{amsmath}
\usepackage{amssymb}
\journalname{Eur. Phys. J. C}

\begin{document}
\title{Rest frames and relativistic effects on de Sitter spacetimes}

\author{Ion I. Cot\u aescu %\inst{1}
\thanks{e-mail: cota@physics.uvt.ro}}
\institute{West University of Timi\c soara, V. P\^ arvan Ave. 4, RO-300223, Timi\c soara, Romania}
\date{Received: date / Revised version: date}

\maketitle
\begin{abstract}
It is shown that the Nachtmann boosting method of introducing  coordinates on de Siter manifolds can be completed with suitable gauge transformations able to keep under control the transformation under isometries of the conserved quantities. With this method, the rest local charts (or natural frames)  are defined pointing out the role of the conserved quantities in investigating the relative geodesic motion. The advantages of this approach can be seen from the applications presented here. For the first time,  the simple kinematic effects, the electromagnetic field of a free falling charge and the binary fission are solved in terms of conserved quantities on the expanding portion of the de Sitter spacetime.  
\end{abstract} %end of abstract
 % end of PACS codes

\section{Introduction}
\label{intro}
In special and general relativity there are two maximally symmetric spacetimes, the Minkowski and  de Sitter  ones \cite{SW}, allowing  translations and, consequently, conserved momenta.  In special relativity the translations play a crucial role in Wigner's theory of the induced representations of the Poincar\' e group \cite{Wig,Mc,WKT} which is  based on the orbital analysis in the energy-momentum space. 

Unfortunately, this method cannot be applied  to the de Sitter spacetimes since here the momentum is combined with other conserved quantities that depend on coordinates and transform among themselves under isometries \cite{C1}. Therefore, in this case we cannot speak about the energy-momentum space and  its orbits.  Nevertheless, despite of this difficulty, we may study how different observers measure the conserved quantities on geodesics, resorting to our previous methods in investigating external symmetries \cite{C1,ES,C2,C3,C4}. Our purpose here is to solve this problem  considering the (conformal) Euclidean and de Sitter-Pailev\' e local charts as inertial natural frames where each geodesic is determined by the conserved momentum  in a certain position at a given moment.  The coordinates of these local charts, as well as the conserved quantities on geodesics, are related among themselves through the de Sitter $SO(1,4)$  isometries that become thus the central pieces of our approach.

The method we use here was proposed initialy by Nachtmann for constructing  covariant representations of the de Sitter isometry group \cite{Nach}.  The idea is to introduce the coordinates of the local charts with the help of  point-dependent  $SO(1,4)$ linear transformations of the embedding space  which are called here {\em boosts}. In this paper we show how the original Nachtmann boosts could be completed with suitable gauge transformations of the Lorentz group, $L^{\uparrow}_{+}$, in order to give rise simultaneously to local coordinates and desired conserved quantities. By using such boosts we define the natural rest frames  of the massive mobiles and derive the  isometry transformations among these frames and other arbitrary ones. We obtain thus the principal original result reported here we call Lorenzian isometries since these play the same role as the Lorentz boosts of special relativity \cite{WKT}.   We upgrade thus the so called de Sitter relativity \cite{dSrel}, opening the door to a large field of applications.  

For marking out the advantages of our approach, we give examples of relativistic effects  that can be solved in terms of conserved quantities on the expanding portion of the de Sitter spacetime.  We start with the elementary relativistic kinematic effects, as the time dilation (or twin paradox) and Lorentz contraction, shoving that these are decreasing in time. The second  example is  the electromagnetic field of a freely falling electric charge with a given momentum.  The last example focuses on the  conserved parameters of the binary fission in arbitrary frames.       

The paper is organized as follows. In the second section we briefly present the de Sitter isometries among the Euclidean and respectively de Sitter-Painlev\' e local charts. The third section is devoted to the classical conserved quantities on timelike geodesics and their transformations under isometries. In the next section we extend the Nachtmann method defining the boosts able to introduce  coordinates giving rise simultaneously to a desired momentum of the moving object in a given point. In the fifth section we apply this method for deriving the Lorenzian isometries. In the next section we discuss the mentioned examples giving the principal technical details while in the last one  we present our concluding remarks.

\section{de Sitter isometries}
\label{sec:2}
Let us start with the de Sitter spacetime $(M,g)$ defined as the hyperboloid of radius $1/\omega$  in the five-dimensional flat spacetime $(M^5,\eta^5)$ of coordinates $z^A$  (labeled by the indices $A,\,B,...= 0,1,2,3,4$) having the metric $\eta^5={\rm diag}(1,-1,-1,-1,-1)$. The local charts $\{x\}$  of coordinates $x^{\mu}$ ($\alpha,\mu,\nu,...=0,1,2,3$) can be introduced on $(M,g)$ giving the set of functions $z^A(x)$ which solve the hyperboloid equation,
\begin{equation}\label{hip}
\eta^5_{AB}z^A(x) z^B(x)=-\frac{1}{\omega^2}\,.
\end{equation}
where  $\omega$ denotes the Hubble de Sitter constant since in our notations  $H$ is reserved for the energy (or Hamiltonian) operator \cite{C1}. 

The de Sitter isometry group is just the gauge group $G(\eta^5)=SO(1,4)$ of the embedding manifold $(M^5,\eta^5)$  that leave  invariant its metric and implicitly Eq. (\ref{hip}). Therefore, given a system of coordinates defined by the functions $z=z(x)$, each transformation ${\frak g}\in SO(1,4)$ defines the isometry $x\to x'=\phi_{\frak g}(x)$ derived from the system of equations $z[\phi_{\frak g}(x)]={\frak g}z(x)$. For studying these isometries we use the canonical parametrization
\begin{equation}
{\frak g}(\xi)=\exp\left(-\frac{i}{2}\,\xi^{AB}{\frak S}_{AB}\right)\in SO(1,4) 
\end{equation}
with skew-symmetric parameters, $\xi^{AB}=-\xi^{BA}$,  and the covariant generators ${\frak S}_{AB}$ of the fundamental representation of the $so(1,4)$ algebra carried by $M^5$. These generators have the matrix elements, 
\begin{equation}
({\frak S}_{AB})^{C\,\cdot}_{\cdot\,D}=i\left(\delta^C_A\, \eta_{BD}^5
-\delta^C_B\, \eta_{AD}^5\right)\,.
\end{equation}
The principal $so(1,4)$ basis-generators with an obvious physical meaning \cite{C1} are the energy ${\frak H}=\omega{\frak S}_{04}$, angular momentum  ${\frak J}_k=\frac{1}{2}\varepsilon_{kij}{\frak S}_{ij}$, Lorentz boosts ${\frak K}_i={\frak S}_{0i}$, and the Runge-Lenz-type vector ${\frak R}_i={\frak S}_{i4}$. In addition, it is convenient to introduce the momentum ${\frak P}_i=-\omega({\frak R}_i+{\frak K}_i)$ and its dual ${\frak Q}_i=\omega({\frak K}_i-{\frak R}_i)$ which are nilpotent matrices (i. e. $({\frak P}_i)^3=({\frak Q}_i)^3=0$) of two Abelian three-dimensional subalgebras, $t(3)_P$ and respectively $t(3)_Q$ generating the Abelian subgroups $T(3)_P$ and $T(3)_Q$ \cite{C2}.  

For understanding the action of the isometries generated by these matrices, we  focus on two principal sets of local charts.  The first one is formed by the {\em conformal} Euclidean charts $\{t_c,\vec{x}_c\}$ which offers us some technical advantages. The conformal time $t_c$ and Cartesian spaces coordinates $x_c^i$ ($i,j,k,...=1,2,3$) are defined by the functions 
\begin{eqnarray}
z^0(x_c)&=&-\frac{1}{2\omega^2 t_c}\left[1-\omega^2({t_c}^2 - \vec{x}_c^2)\right]\,,
\nonumber\\
z^i(x_c)&=&-\frac{1}{\omega t}x_c^i \,, \label{Zx}\\
z^4(x_c)&=&-\frac{1}{2\omega^2 t_c}\left[1+\omega^2({t_c}^2 - \vec{x}_c^2)\right]\,,
\nonumber
\end{eqnarray}
written with the vector notation, $\vec{x}=(x^1,x^2,x^3)\in {\Bbb R}^3\subset M^5$. These charts  cover the expanding part of $M$ for $t_c \in (-\infty,0)$
and $\vec{x}_c\in {\Bbb R}^3$ while the collapsing part is covered by
similar charts with $t_c >0$. In both these cases we have the same conformal flat line element,
\begin{equation}\label{mconf}
ds^{2}=\eta^5_{AB}dz^A(x_c)dz^B(x_c)=\frac{1}{\omega^2 {t_c}^2}\left({dt_c}^{2}-d\vec{x}_c^2\right)\,.
\end{equation}
We stress that here we restrict ourselves to consider only the expanding portion which is a possible  model of our expanding universe.

Another choice are the de Sitter-Painlev\' e coordinates $\{t, \vec{x}\}$ on the same portion that can be introduced  directly substituting
\begin{equation}\label{EdS}
t_c=-\frac{1}{\omega}e^{-\omega t}\,, \quad \vec{x}_c=\vec{x}e^{-\omega t}\,,
\end{equation}
where $t\in(-\infty, \infty)$ is the {\em proper} time while $x^i$ are the 'physical' Cartesian space coordinates. Then the line element reads
\begin{equation}
ds^2=(1-\omega^2 {\vec{x}}^2)dt^2+2\omega \vec{x}\cdot d\vec{x}\,dt -d\vec{x}\cdot d\vec{x}\,. 
\end{equation}
Notice that this chart is useful in applications since in the flat limit (when $\omega \to 0$) its coordinates become just the Cartesian ones of the Minkowski spacetime.  

Now we can briefly review the effects of the isometries $x_c\to x_c'=\phi_{\frak g}(x_c)$ of the Euclidean chart \cite{C2} or of the corresponding ones, $x\to x'=\phi_{\frak g}(x)$, in the de Sitter- Painlev\' e coordinates.  We observe first that the transformations ${\frak g}\in SO(3)\subset SO(4,1)$  generated by ${\frak J}_i$ are simple rotations of $z^i$ as well as of the Cartesian coordinates  $x_c^i$ and $x^i$ which transform alike since this symmetry is global. The transformations  ${\frak g}={\rm exp}(-i\alpha {\frak H})$, with $\alpha=\omega\xi$, produce the dilation of the conformal coordinates, 
$t_c\to t_c\,e^{\alpha}$ and $x_c^i\to x_c^i e^{\alpha}$, which appear in the chart $\{t,\vec{x}\}$ as simple time translations $t\to t-\xi$ at fixed $\vec{x}$. For this reason we denote this subgroup with $T(1)_H$. The transformations  of the Abelian subgroup $T(3)_P$  give rise to the space translations at fixed time. More  interesting are the $T(3)_Q$ transformations generated by ${\frak Q}_i/$, which produce more complicated isometries of the Euclidean charts \cite{C2} that can be rewritten in the de Sitter- Painlev\' e ones by using  the substitution (\ref{EdS}).

\section{Classical conserved quantities}
\label{sec:3}

The classical conserved quantities under de Sitter isometries  can be calculated with the help of  the  Killing vectors  $k_{(AB)}$ of the de Sitter manifold $(M,g)$ \cite{ES,C1}. According to the general definition of the Killing vectors in the pseudo-Euclidean spacetime $(M^5,\eta^5)$, we may consider the following identity
\begin{equation}
K^{(AB)}_Cdz^C=z^Adz^B-z^Bdz^A=k^{(AB)}_{\mu}dx^{\mu}\,,
\end{equation}
giving the covariant components  of the Killing vectors in an arbitrary chart $\{x\}$ of  $(M,g)$ as
\begin{equation}\label{KIL}
k_{(AB)\,\mu}=\eta^5_{AC}\eta^5_{BD}k^{(CD)}_{\mu}= z_A\partial_{\mu}z_B-z_B\partial_{\mu}z_A\,, 
\end{equation}
where $z_A=\eta_{AB}z^B$. 

The principal conserved quantities along a timelike geodesic of a point-like particle of mass $m$ and momentum $\vec{P}$ \cite{C1} have the general form  ${\cal K}_{(AB)}(x,\vec{P})=\omega k_{(AB)\,\mu}m u^{\mu}$ where $u^{\mu}=\frac{dx^{\mu}(s)}{ds}$ are the components of the covariant four-velocity that satisfy  $u^2=g_{\mu\nu}u^{\mu}u^{\nu}=1$. The conserved quantities with physical meaning \cite{C1} are, 
\begin{eqnarray}
{\frak H} &\to& E=\omega  k_{(04)\,\mu}m u^{\mu}\label{consE}\\
{\frak J}_i&\to& L_i= \frac{1}{2}\varepsilon_{ijk}k_{(jk)\,\mu}m u^{\mu}\\
{\frak K}_i&\to& K_i=  k_{(0i)\,\mu} m u^{\mu}\\
{\frak R}_i&\to& R_i=  k_{(i4)\,\mu} m u^{\mu}\,,\label{consR}
\end{eqnarray}
where $E$ is the conserved energy,  $L_i$ are the usual components of angular momentum while  $K_i$ and $R_i$  are related to  the conserved momentum, $P^i=-\omega(R_i+K_i)$,  and  its dual, $Q^i=\omega(K_i-R_i)$ \cite{C1}. Thus we can construct the five-dimensional matrix 
\begin{equation}
{\cal K}(x,\vec{P})=
\left(
\begin{array}{ccccc}
0&\omega K_1&\omega K_2&\omega K_3&E\\
-\omega K_1&0&\omega L_3&-\omega L_2&\omega R_1\\
-\omega K_2&-\omega L_3&0&\omega L_1&\omega R_2\\
-\omega K_3&\omega L_2&-\omega L_1&0&\omega R_3\\
-E&\omega R_1&-\omega R_2&-\omega R_3&0
\end{array}\right)\,,\label{KK}
\end{equation}
whose elements  transform  as  a five-dimensional skew-symmetric tensor on $M^5$, according to the rule 
\begin{equation}\label{KAB}
{\cal K}_{(AB)}(x',{\vec{P}}')={\frak g}_{A\,\cdot}^{\cdot\,C}\,{\frak g}_{B\,\cdot}^{\cdot\,D}\,{\cal K}_{(CD)}(x,\vec{P})\,,
\end{equation}
for all ${\frak g}\in SO(1,4)$. Here  ${\frak g}_{A\,\cdot}^{\cdot\,B}=\eta^5_{AC}\,{\frak g}^{C\,\cdot}_{\cdot \,D}\, \eta^{5\,BD}$ are the matrix elements of the adjoint matrix $\overline{\frak g}=\eta^5\,{\frak g}\,\eta^5$. Thus,  Eq. (\ref{KAB}) can be written as  ${\cal K}(x',{\vec{P}}')=\overline{\frak g}\,{\cal K}(x,\vec{P})\,\overline{\frak g}^T$ or simpler, ${\cal K}'=\overline{\frak g}\,{\cal K}\,\overline{\frak g}^T$. 

The  properties of the above defined conserved quantities may be studied by choosing a convenient local chart.  Technically speaking the best choice is that of the Euclidean  charts $\{t_c,\vec{x}_c\}$ where the contravariant components of the Killing vectors can be calculated according to Eq. (\ref{KIL}) as 
\begin{eqnarray}
&&k^0_{(0i)}=k^0_{(4i)}=-\omega t_c x_c^i\,,\nonumber\\
&&k^j_{(0i)}=k^j_{(4i)}+\frac{1}{\omega}\,\delta^j_i=- \omega x_c^i x_c^j+\delta^j_i\chi\,,\nonumber\\
&&k^0_{(ij)}=0\,,\quad k^l_{(ij)}=\delta^l_i x_c^j-\delta^l_j
x_c^i\,,\label{chichi}\\
&& k^0_{(04)}=-t\,,\quad k^i_{(04)}=-x_c^i\,.\nonumber
\end{eqnarray}
where we denote 
\begin{equation}
\chi=\frac{1}{2\omega}\left[1-\omega^2(t_c^2-{\vec{x}_c}^2)\right]\,.
\end{equation}
Taking into account that the particle of mass $m$ has the momentum $\vec{P}$  of components 
$P^i=-\omega m (k_{(0i)\,\mu} + k_{(i4)\,\mu})u_c^{\mu}$, 
we find the components of the four-velocity, 
\begin{eqnarray}
u_c^0&=&\frac{dt_c}{ds}=-\omega t_c \sqrt{1+\frac{\omega^2{P}^{2}}{m^2}t_c^2}\,,\nonumber\\
u_c^i&=&\frac{d{x_c^i}}{ds}=(\omega t_c)^2 \frac{P^i}{m}\,,\label{uu}
\end{eqnarray}
where  we denote $P=|{\vec{P}}\,|$. Hereby we obtain the geodesic trajectory \cite{C1},
\begin{eqnarray}
{x_c}^i(t_c)&=&{x_c}_0^i+\frac{P^i}{\omega {P}^
2}\nonumber\\
&\times& \left(\sqrt{m^2+{P}^{2}\omega^2 {t_{c0}}^2}-\sqrt{ m^2+{P}^2
\omega^2 t_c^2}\, \right)\,,\label{geodE}
\end{eqnarray}
of a massive particle passing through the space point $\vec{x}_{c0}$ at time ${t_{c0}}$. This  is completely determined by the initial condition  $\vec{x}_c({t_{c0}}) =\vec{x}_{c0}$  and the conserved momentum $\vec{P}$. Therefore, the  conserved quantities  in an arbitrary point $(t_c,\vec{x}_c(t_c))$ of the geodesics  depend only on this point and the momentum $\vec{P}$. Then, by substituting the components (\ref{chichi}) and (\ref{uu}) in Eqs. (\ref{consE})-(\ref{consR}) we find  the other conserved quantities  \cite{C1},
\begin{eqnarray}
E&=&\omega\, \vec{x}_c(t_c)\cdot \vec{P}+\sqrt{ m^2+{P}^{2}\omega^2 t_c^2}\,,\label{Ene}\\
L_i&=&\varepsilon_{ijk} x^j_c(t_c) P^k\,,\\
Q^i&=&2\omega x_c^i(t_c)E+\omega^2P^i[t_c^2-\vec{x}_c(t_c)^2]\,.\label{Q}
\end{eqnarray}
that satisfy the  obvious identity
\begin{equation}
E^2-\omega^2 {\vec{L}}^2-\vec{P}\cdot \vec{Q}=m^2
\end{equation}
corresponds to the first Casimir operator of the $so(1,4)$ algebra \cite{C1}. In the flat limit, $\omega\to 0$ when $-\omega t_c\to 1$ and $\vec{Q} \to \vec{P}$ this identity  becomes just  the usual mass-shell condition $p^2=m^2$ of special relativity.   

An important particular case is of the rest particle with $\vec{P}=0$ staying  in an arbitrary point $(t_c, \vec{x}_c)$  on a world line along the vector field $-\omega t_c \partial_{t_c}$. Then the rest energy, $E=m$, is the same as in special relativity, $\vec{L}=0$ and $\vec{Q}=2\omega m \vec{x}_c$ such that the matrix
\begin{equation}
{\cal K}(x_c,0)=\left(
\begin{array}{ccccc}
0&m\omega x^1_c&m\omega x^2_c&m\omega x^3_c&m\\
-m\omega x^1_c&0&0&0&-m\omega x^1_c\\
-m\omega x^2_c&0&0&0&-m\omega x^2_c\\
-m\omega x^3_c&0&0&0&-m\omega x^3_c\\
-m&m\omega x^1_c&m\omega x^2_c&m\omega x^3_c&0
\end{array}\right)\,,\nonumber
\end{equation}
is independent on $t_c$. If we suppose, in addition, that the  particle  stays at rest in origin,  $\vec{x}_c=0$,  then this matrix takes the simplest form 
\begin{equation}
{\cal K}_o=\left(
\begin{array}{ccccc}
0&0&0&0&m\\
0&0&0&0&0\\
0&0&0&0&0\\
0&0&0&0&0\\
-m&0&0&0&0
\end{array}\right)\,,
\end{equation}
depending only on the particle mass. 

We observe that the stable group of the matrix ${\cal K}_o$ is the group $SO(3)\otimes T(1)_H \subset SO(1,4)$ since $\overline{\frak g}\,{\cal K}_{o}\,\overline{\frak g}^T= {\cal K}_{o}$ for any transformation ${\frak g}$ of this group. On the other hand, the $T(3)_P$ translations ${\frak g}=\exp(-i\vec{\xi}\cdot {\frak P})$ have the action $\overline{\frak g}\,{\cal K}_{o}\,\overline{\frak g}^T= {\cal K}(\vec{\xi},0)$  without giving rise to the momentum components. Thus we conclude that all the isometries of the subgroup $G_o=SO(3)\otimes T(1)_H\otimes T(3)_P$ preserve the rest states with $\vec{P}=0$ changing only the positions of the particle at rest. This group plays here the same role as the little group of the orbit $p^2=m^2$ in the energy-momentum space of special relativity.

\section{Boosting coordinates and momenta}
\label{sec:4}

Many years ago Nachtmann proposed a boosting method for deriving covariant reprentations of the de Sitter isometry group  induced by the gauge group $G(\eta)=L^{\uparrow}_{+}$ that leave invariant the metric $\eta={\rm diag}(1,-1,-1,-1)$ of the Minkowskian pseudo-Euclidean model of $(M,g)$ \cite{Nach}. This method offers new techncal opportunities that allowed us to derive the generalized Rindler transformation on the de Sitter manifolds \cite{C3}. This encourages us to apply the same technique  for solving the problem of the classical relative geodesic motion on the expanding portion of the de Sitter spacetime. In what follows we develop our formalism  denoting for brevity $G\equiv G(\eta)=L^{\uparrow}_{+}$ and $G_5\equiv G(\eta^5)=SO(1,4)$.  

The Nachtmann method uses the Wigner orbital analysis  but in configurations instead of momentum representation \cite{Nach}. This can be done since the de Sitter manifold is isomorphic with the space of left cosets $G_5/G$. Indeed, if one fixes the point $z_o=(0,0,0,0,\omega^{-1})^T$ in $M^5$ (of local Euclidean coordinates $(-\omega^{-1},0,0,0)$ or $(0,0,0,0)$ de Sitter-Painlev\' e ones) then the whole de Sitter manifold can be built as the orbit $M=\{{\frak g} z_o |{\frak g}\in G_5/G\} \subset M^5$ since the subgroup $G$ is just the stable group of $z_o$ (obeying ${\frak g}z_o=z_o\,,\, \forall {\frak g}\in G$). Then, any point $z(x)\in M$ can be reached  applying the boost, ${\frak b}(x):z_o\to z(x)={\frak b}(x)z_o$, defining the functions $z^A(x)$ of the local coordinates $\{x\}$. In fact, these boosts are sections in the principal fiber bundle on  $(M,g)\sim  G_5/G$ whose fiber is just the isometry group $G_5$.   

This formalism offers one the advantage of defining  the {\em canonical} five-dimensional 1-forms associated to the boost ${\frak b}(x)$ that read \cite{Nach}
\begin{equation}
\hat\omega(x)={\frak b}^{-1}(x)d\,{\frak b}(x)\,z_o\,.
\end{equation}  
The components 
\begin{equation}
\hat\omega^{\hat\alpha}(x)=\hat e^{\hat\alpha}_{\mu}(x)dx^{\mu}\,,\quad 
\hat\omega^{4}(x)=0\,,
\end{equation} 
are labeled by the local indices $\hat\alpha,\hat\mu,...$, with the same range as the natural ones. These define the canonical gauge fields (or tetrads)  $\hat e^{\hat\mu}$ of the local co-frames associated to the fields $e_{\hat\mu}$  of the orthogonal local frames \cite{Nach}.  In general, the boosts are defined up to an {\em arbitrary} gauge, ${\frak b}(x)\to {\frak b}(x)\lambda^{-1}(x)$, $\lambda(x)\in G$, that does not affect the functions $z^A(x)$ but changes the gauge fields  transforming the  1-forms  as $\hat\omega(x) \to \lambda(x)\,\hat\omega(x)$ \cite{Nach,C2}.    

                                                                                                                                                                                                                                                                                           The structure of the boost transformation determines the type of the chart that has to be defined in this manner. The Euclidean chart $\{t_c,\vec{x}_c\}$ under consideration here is boosted by the transformation \cite{Nach},
\begin{equation}
{\frak b}(t_c,\vec{x}_c)=\exp(-ix^i {\frak P}_i) \exp(-i\alpha {\frak H})\,,    
\end{equation}
which depends on  $\alpha=\ln(-\omega t_c)$ having the form 
\begin{eqnarray}\label{boost}
&&{\frak b}(t_c,\vec{x}_c)=\nonumber\\
&&-\left(
\begin{array}{ccccc}
\frac{1+\omega^2(t_c^2+\vec{x}_c^2)}{2\omega t_c}&-\omega x_c^1&-\omega x_c^2&-\omega x_c^3&-\frac{1-\omega^2(t_c^2-\vec{x}_c^2)}{2\omega t_c}\\
\frac{x_c^1}{t_c}&1&0&0&\frac{x_c^1}{t_c}\\
\frac{x_c^2}{t_c}&0&1&0&\frac{x_c^2}{t_c}\\
\frac{x_c^3}{t_c}&0&0&1&\frac{x_c^3}{t_c}\\
\frac{1-\omega^2(t_c^2+\vec{x}_c^2)}{2\omega t_c}&\omega x_c^1&\omega x_c^2&\omega x_c^3&\frac{1+\omega^2(t_c^2-\vec{x}_c^2)}{2\omega t_c}\\
\end{array}\right)\,.\nonumber\\\label{boo}
\end{eqnarray}
It is worth observing that for $t_c=-\omega^{-1}$ and $\vec{x}_c=0$ we obtain the identity transformation,  ${\frak b}(\omega^{-1},0)={\frak e}$, since these are just the coordinates defining the fixed point $z_o$. 

Assuming now that a particle stays at rest in $z_o$ having $\vec{P}=0$ and the conserved quantities given by the matrix ${\cal K}_o$, we observe that the boost (\ref{boo}) is not able to give rise to momentum components since this is a transformation of the subgroup $G_o$ whose isometries preserve the rest states.  This means that for boosting momenta we need more.  The unique possibility is to look for a suitable gauge transformation  giving rise to a non-vanishing momentum. The solution we propose here is to construct the new boost 
\begin{equation}
{\frak b}(t_c,\vec{x}_c,\vec{P})={\frak b}(t_c,\vec{x}_c) {\frak l}(t_c,\vec{P})
\end{equation}
where the matrix 
\begin{eqnarray}
&&{\frak l}(t_c,\vec{P})=\exp\left(i \vec{P}\cdot \vec{\frak K}\,\frac{1}{P}{\rm arcsinh}\,\frac{\omega P t_c}{m}\right)=\nonumber\\
&&\left(
\begin{array}{ccccc}
\frac{E_{t_c}}{m}&-\frac{\omega t_c P^1}{m}&-\frac{\omega t_c P^2}{m}&-\frac{\omega t_c P^3}{m}&0\\
-\frac{\omega t_c P^1}{m}&1+{n_p^1}^2\nu_{t_c}&n_p^1 n_p^2\nu_{t_c}&n_p^1 n_p^3\nu_{t_c}&0\\
-\frac{\omega {t_c} P^2}{m}&n_p^1 n_p^2\nu_{t_c}&1+{n_p^2}^2\nu_{t_c}&n_p^2 n_p^3\nu_{t_c}&0\\
-\frac{\omega {t_c} P^3}{m}&n_p^1 n_p^3\nu_{t_c}&n_p^2 n_p^3\nu_{t_c}&1+{n_p^3}^2\nu_{t_c}&0\\
0&0&0&0&1
\end{array}\right),
\end{eqnarray}
is a time-dependent Lorentz boost of the gauge group $G$ written with the notations $E_{t_c}=\sqrt{m^2+\omega^2 P^2 t_c^2}$, $\nu_{t_c}=\left(\frac{E_{t_c}}{m}-1\right)$ and $n^i_P=\frac{P^i}{P}$.   Notice that  this boost is defined up to an arbitrary point-dependent rotation (${\frak l}\to {\frak l}{\frak r}(x_c))$ since the $SO(3)$ group  is a stable group for both $z_o$ and  ${\cal K}_o$.   

Concluding we may say that the principal new result obtained in this section is the boost ${\frak b}(t_c,\vec{x}_c,\vec{P})$ that  brings the particle of mass $m$ from $z_o$ to the point $x_c$ defined by $z(x_c)={\frak b}(t_c,\vec{x}_c,\vec{P}) z_o={\frak b}(t_c,\vec{x}_c) z_o$,  where this particle gets the momentum $\vec{P}$ determining the conserved quantities encapsulated in the matrix 
\begin{equation}
{\cal K}(t_c,\vec{x}_c,\vec{P})=\overline{\frak b}(t_c,\vec{x}_c,\vec{P})\,{\cal K}_o\,\overline{\frak b}(t_c,\vec{x}_c,\vec{P})^T
\end{equation} 
that has the form  (\ref{KK}) with components given by Eqs.  (\ref{Ene}) and (\ref{Q}).  

\section{Rest frames and Lorenzian isometries}
\label{sec:5}

The problem of relative motion in special or general relativity is to establish how  the observers  in different local charts related through isometries measure the same geodesic motion.   In special relativity this problem is solved considering  inertial frames transforming under Poincar\' e isometries. Similar isometries in the more complicated case of the de Sitter spacetime have to be derived applying the above boosting method to the Euclidean local charts or de Sitter-Painlev\' e ones. 

Let us consider first  two Euclidean charts $\{t_c,\vec{x}_c\}$ and $\{t_c',\vec{x}'_c\}$ in which the observers $O$ and respectively $O'$ measure the same geodesic motion of a  particle of mass $m$. Then, assuming that the observer $O$ measures the parameters $({t_{c0}},\vec{x}_{c0},\vec{P})$ while $O'$ observes other parameters, $(t'_{c0},\vec{x}'_{c0},\vec{P}')$, of the same particle, we may deduce the isometry  relating these charts. Indeed, starting with  $z({t_{c0}},\vec{x}_{c0})={\frak b}({t_{c0}},\vec{x}_{c0},\vec{P})z_o$ and $z(t'_{c0},\vec{x}'_{c0})={\frak b}(t'_{c0},\vec{x}'_{c0},\vec{P}')z_o$ we observe that the coordinates of these charts are related through the isometry  $x_c=\phi_{{\frak g}_*}(x_c')$ generated by the transformation  
\begin{eqnarray}\label{rel}
{\frak g}_*&=&{\frak b}(t_{c0},\vec{x}_{c0},\vec{P}){\frak b}({t'_0},\vec{x}'_{c0},\vec{P}')^{-1}\nonumber\\
&=&{\frak b}(t_{c0},\vec{x}_{c0}) {\frak l}(t_{c0},\vec{P})\,{\frak r}\,{\frak l}(t'_{c0},\vec{P}')^{-1}{\frak b}(t'_{c0},\vec{x}'_{c0})^{-1}.
\end{eqnarray}
As mentioned, this is defined up to an arbitrary rotation ${\frak r}\in SO(3)$ which is fixed here to the unit isometry, ${\frak r}={\frak e}$. The conserved quantities that can be observed by $O$ and $O'$ are  related through 
\begin{equation}
{\cal K}(t_c,\vec{x}_c,\vec{P})=\overline{\frak g}_*\,\,{\cal K}(t_c',\vec{x}'_c,\vec{P}')\overline{\frak g}_*^T\,.
\end{equation}
 
An useful  application is the definition of the natural rest frames of a particle of mass $m$ having a given momentum $\vec{P}$. Let us assume that the chart $\{t_c, \vec{x}_c\}$ is the frame of the fixed observer $O$ while the observer $O'$ moves with the mobile chart $\{t'_c, \vec{x}'_c\}$ in which the particle of mass $m$ stays at rest (with $\vec{P}'=0$) in  $\vec{x}'_c=\vec{x}'_{c0}$,  having the world line along the vector field $-\omega t'_c\partial_{t'_c}$. In general, the clocks of these frames are not synchronized such that the fixed observer $O$ may measure the  parameters $(t_{c0},\vec{x}_{c0}, \vec{P})$ corresponding to another initial condition. Then, by calculating explicitly the matrix (\ref{rel}) in this case, ${\frak g}_*={\frak b}((t_{c0},\vec{x}_{c0},\vec{P}){\frak b}({t'_{c0}},\vec{x}'_{c0},0)^{-1}$,  we find the general isometry transformation given in the Appendix A which suggests us to synchronize the clocks by choosing the following suitable initial conditions
\begin{equation}\label{incon}
\vec{x}'_{c0}=\vec{x}_{c0}=0\,, \quad t'_{c0}=t_{c0}=-\frac{1}{\omega}\,.
\end{equation}
This means that the particle of $m$ stays in the origin of the rest frame $O'$ that is passing through the origin of the fixed frame $O$ at  the initial time $t_c'=t_c=-\omega^{-1}$. The advantage of these initial conditions is that ${\frak b}(-\omega^{-1},0)={\frak e}$ and the energy takes the same form as in the flat case,  $E=\sqrt{m^2 +P^2}$. Consequently, the  transformation (\ref{rel}), denoted from now by ${\frak g}_*={\frak g}(\vec{P})$, becomes     
\begin{eqnarray}
{\frak g}(\vec{P})&=&{\frak l}(-\omega^{-1},\vec{P})=\exp\left(-i \vec{P}\cdot \vec{\frak K}\,\frac{1}{P}\,{\rm arcsinh}\frac{P}{m}\right)\nonumber\\
&&=
\left(
\begin{array}{ccccc}
\frac{E}{m}&\frac{P^1}{m}&\frac{ P^2}{m}&\frac{ P^3}{m}&0\\
\frac{P^1}{m}&1+{n_p^1}^2\nu&n_p^1 n_p^2\nu&n_p^1 n_p^3\nu&0\\
\frac{P^2}{m}&n_p^1 n_p^2\nu&1+{n_p^2}^2\nu&n_p^2 n_p^3\nu&0\\
\frac{ P^3}{m}&n_p^1 n_p^3\nu&n_p^2 n_p^3\nu&1+{n_p^3}^2\nu&0\\
0&0&0&0&1
\end{array}\right)\,,\label{Lorbust}
\end{eqnarray}
where now we denote $\nu=\left(\frac{E}{m}-1\right)$. The four-dimensional restriction of this transformation is a genuine Lorentz boost such that  
 ${\frak g}(\vec{P})^{-1}={\frak g}(-\vec{P})$ and ${\frak g}(0)={\frak e}$.

Hereby we obtain the principal new result of this paper, namely the Lorenzian  isometry  $x_c=\phi_{{\frak g}(\vec{P})}(x_c')$ between the rest frames $O'$ and that of the fixed observer $O$  that reads 
\begin{eqnarray}
t_c(t_c',\vec{x}'_c)&=&\frac{t'_c}{\Delta_c'}\,,\label{Eq1}\\
{\vec{x}_c}(t_c',\vec{x}'_c)&=&\frac{1}{\Delta_c'}\left\{\vec{x}'_c+\frac{\vec{P}}{m}\left[\frac{\vec{x}'_c\cdot\vec{P}}{E+m}\right.\right.\nonumber\\
&&\left.\left.+\frac{1}{2\omega}\left(1-\omega^2({t_c'}^2-{\vec{x}'_c}^2)\right)\right]\right\}\,,\label{Eq2}
\end{eqnarray}
where
\begin{equation}
\Delta_c'=1+\frac{\omega}{m}\, \vec{x}'_c\cdot\vec{P}+\frac{E-m}{2m}\left(1-\omega^2({t_c'}^2-{\vec{x}'_c}^2)\right)\,.
\end{equation}
In addition, we can write  the transformation rule 
\begin{equation}\label{KKg}
{\cal K}(t_c,\vec{x}_c,\vec{P}_1)=\overline{\frak g}(\vec{P})\,{\cal K}(t'_c,\vec{x}'_c,\vec{P}'_1)\,\overline{\frak g}(\vec{P})^T\,,
\end{equation}
among the conserved quantities of an arbitrary geodesic with the parameters $(t_c,\vec{x}_c,{\vec{P}_1})$  observed by $O$ and $(t_c',\vec{x}'_{c},\vec{P}'_1)$ observed by $O'$. The inverse Lorenzian isometry, $x'_c=\phi_{{\frak g}(-\vec{P})}(x_c)$, may be obtained by changing $x_c$ and  $x'_c$ between themselves and $\vec{P} \to -\vec{P}$.

The Lorenzian isometries may be written in de Sitter-Painlev\' e coordinates denoting  by $\{t,\vec{x}\}$  those of the fixed frame  and by $\{t', \vec{x}'\}$  the rest frame ones. We keep the initial conditions (\ref{incon}) that take now the natural form
\begin{equation}\label{icon1}
t_0=t'_0=0\,, \quad \vec{x}_0={\vec{x}_0}^{\,\prime}=0\,.
\end{equation}
Then, the Lorenzian isometry $x=\phi_{{\frak g}(\vec{P})}(x')$ reads
\begin{eqnarray}
t(t',{\vec{x}}')&=&\frac{1}{\omega}\, \ln\left( e^{\omega t'}+\frac{\omega}{m}{\vec{x}}'\cdot \vec{P} + \frac{E-m}{m}\,\omega\Theta'  \right)\,,\label{Eq1A}\\
{\vec{x}}(t',{\vec{x}}')&=&{\vec{x}}'+\frac{\vec{P}}{m}\left(\frac{{\vec{x}}'\cdot\vec{P}}{E+m}+\Theta'\right)\,,\label{Eq2A}
\end{eqnarray}
where we used again the identity $m^2+ P^2=E^2$ and denote 
\begin{equation}
\Theta'=\frac{1}{2\omega}\left[e^{\omega t'}-e^{-\omega t'}(1-\omega^2 {{\vec{x}}'}^2)\right]\,.
\end{equation}
Since these transformations have to be used in applications,  we write explicitly the inverse Lorenzian isometry $x'=\phi_{{\frak g}(-\vec{P})}(x)$ which has the transformation rules
\begin{eqnarray}
t'(t,{\vec{x}})&=&\frac{1}{\omega}\, \ln\left( e^{\omega t}-\frac{\omega}{m}{\vec{x}}\cdot \vec{P} + \frac{E-m}{m}\,\omega\Theta  \right)\,,\label{Eq1B}\\
{\vec{x}}'(t,{\vec{x}})&=&{\vec{x}}+\frac{\vec{P}}{m}\left(\frac{{\vec{x}}\cdot\vec{P}}{E+m}-\Theta\right)\,,\label{Eq2B}
\end{eqnarray}
where now we denote  
\begin{equation}
\Theta=\frac{1}{2\omega}\left[e^{\omega t}-e^{-\omega t}(1-\omega^2 {{\vec{x}}}^2)\right]\,.
\end{equation}

We must specify that there are relativistic problems which do not depend on the mass of the  particle carrying the mobile frame. Then we can eliminate this mass changing the parametrization of the Lorenzian isometries in terms of the conserved velocity $\vec{V}=\frac{\vec{P}}{E}$ and denoting  
\begin{equation}
\frac{E}{m}=\gamma\,,\quad \frac{P}{m}=\gamma V\,,\quad  \gamma=\frac{1}{\sqrt{1-V^2}}\,.
\end{equation}  
It is not difficult to show that $\vec{V}$ is the velocity of the origin $O'$ of the mobile frame when this is passing though the origin $O$ of the fixed frame at time $t=t'=0$. This velocity is the same in the charts with Euclidean or de Sitter-Pailev\' e coordinates since we adopted convenient initial conditions. 

For small values of $\omega$ we may consider the expansions 
\begin{eqnarray}
t&=&\frac{E}{m}\,t' +\frac{{\vec{x}}'\cdot \vec{P}}{m}+\frac{1}{2m^2}\left[m(E-m)
{{\vec{x}}'}^2\right. \nonumber\\
&&\hspace*{24mm}\left.-(Et'+\vec{P}\cdot{\vec{x}}')^2\right]\omega +{\cal O}(\omega^2)\,,\\
{\vec{x}}&=&{\vec{x}}'+\frac{\vec{P}}{m}\left[\frac{{\vec{x}}'\cdot\vec{P}}{E+m}+t' \right]
+\frac{1}{2m}{{\vec{x}}'}^2 \vec{P}\, \omega +{\cal O}(\omega^2)\,,
\end{eqnarray}
instead of Eqs. (\ref{Eq1A}) and (\ref{Eq2A}). We obtain thus the corrections of the first order and verify that for  $\omega=0$ we recover  just the usual Lorentz transformations between the rest frame of a particle of mass $m$ of momentum $\vec{P}$ and the frame of the fixed observer  in  Minkowski spacetime.

\section{Relativistic effects}
\label{sec:6}

The isometries studied here my be used in various applications from the elementary relativistic effects up to the study of the properties of the covariant fields. In what follows we give few simple examples that can be studied in this framework. 

\subsection{Kinematic effects}
\label{sec:6.1}

As mentioned, the motion of a particle of mass $m$ is completely determined by the conserved momentum $\vec{P}$ and the initial condition. The observer $O$ measure the kinematic parameters of  this particle whose  geodesic trajectory in the Euclidean chart is given by Eq. (\ref{geodE}) complying with  initial conditions (\ref{incon}). The covariant four-velocity remains of the form (\ref{uu}) since this depends only on $\vec{P}$ regardless the initial conditions.  We note that apart from $\vec{P}$ and $E$ the other conserved quantities are less relevant since in this case we have  $\vec{L}=0$ and $\vec{Q}=\vec{P}$. 

Now we focus on the motion of the particle of mass $m$ in the local chart $O$ with de Sitter-Painlev\' e coordinates for which we use the initial conditions  (\ref{icon1}). In order to avoid confusion we denote the coordinates of this particle in the frame $O$ by $t_*$ and $\vec{x}_*$ and write the geodesic equation 
\begin{equation}\label{geodP}
\vec{x}_*(t_*)=\frac{\vec{P}}{\omega {P}^2}\,\left(Ee^{\omega t_*}-\sqrt{ m^2 e^{2\omega t_*}+{P}^2}\, \right)\,,
\end{equation}
resulted straightforwardly from Eqs. (\ref{EdS}) and (\ref{geodE}). The covariant four-velocity  can be derived from Eqs. (\ref{EdS}) and (\ref{uu}) as
\begin{eqnarray}
u^0_*&=&\frac{dt_*}{ds}=\sqrt{1+\frac{P^2}{m^2}\,e^{-2\omega t_*}}\,,\\
u^i_*&=&\frac{dx^i_*}{ds}=\frac{P^i}{m}e^{-\omega t_*}+\omega x^i_* (t_*)\sqrt{1+\frac{P^2}{m^2}\,e^{-2\omega t_*}}\,,
\end{eqnarray}
laying out the relation between the covariant momentum $p^{\mu}=mu^{\mu}$ and the conserved one. Hereby we observe that for large values of $t_*$ we recover the well-known Hubble law since then $u^0_* \to 1$ and $u^i_* \to \omega x_*^i(t_*)$ where   $\vec{x}_*(t_*) \to \vec{P}[\omega(E+m)]^{-1} e^{\omega t_*}$\,.
  
However, a good test is to obtain the geodesic equation (\ref{geodP}) exploiting the isometry transformations (\ref{Eq1A}) and (\ref{Eq2A}) that give the coordinates of the particle of mass $m$ as
\begin{eqnarray}
t_*&=&t(t',0)=\frac{1}{\omega}\, \ln\left( e^{\omega t'} + \frac{E-m}{m}\,\sinh(\omega t')  \right)\,,\label{Eq1C}\\
\vec{x}_*&=&{\vec{x}}(t',0)=\frac{\vec{P}}{m\omega}\sinh(\omega t')\,.\label{Eq2C}
\end{eqnarray}
Solving the first equation we find the function $t'(t_*)$ of the form,
\begin{equation}\label{ttst}
t'=\frac{1}{\omega}\ln\left[\frac{me^{\omega t_*}+\sqrt{m^2 e^{2\omega t_*}+P^2} }{E+m}\right]\,,
\end{equation}
that substituted in Eq. (\ref{Eq2C}) leads just to the geodesic equation (\ref{geodP}).
  
Other interesting applications are the time dilation (observed in the so called twin paradox) and the Lorentz contraction which in this case are quite complicated since the these effects are strongly dependent on the position where the time and length are measured. Therefore, for giving a mere simple example, we assume  that the measurements are performed in a small neighborhood  of the carrier particle (where ${\vec{x}}'=0$).   Here we consider the general relations 
\begin{eqnarray}
\delta t &=&\left.\frac{\partial t(t',{\vec{x}}')}{\partial t'}\right| _{{\vec{x}}'=0}\delta t' + \left.\frac{\partial t(t',{\vec{x}}')}{\partial x^{\prime\, i}}\right| _{{\vec{x}}'=0}\delta  x^{\prime\, i}\,,\\
\delta x^j &=&\left.\frac{\partial x^j(t',{\vec{x}}')}{\partial t'}\right| _{{\vec{x}}'=0}\delta t' +\left.\frac{\partial x^j(t',{\vec{x}}')}{\partial x^{\prime\, i}}\right| _{{\vec{x}}'=0}\delta  x^{\prime\, i}\,,
\end{eqnarray} 
among the quantities $\delta t, \delta x^j$ and $\delta t',\delta x^{\prime\, j}$ supposed to be measured by the observers $O$ and respectively $O'$.  First we consider a clock in $m$ indicating $\delta t'$ without changing its position such that $\delta x^{\prime\, i}=0$. Then from  Eq. (\ref{ttst}), after a little calculation, we obtain the time dilation observed by $O$,    
\begin{equation}
\delta t=\delta t' \left(1+\frac{P^2}{m^2}e^{-2\omega t_*}\right)^{\frac{1}{2}}\,,
\end{equation}   
Similarly but with the supplemental simultaneity condition $\delta t=0$ we derive the Lorentz contraction along the direction of $\vec{P}$ that reads
\begin{equation}
\delta x_{||}=\delta x'_{||}\left(1+\frac{P^2}{m^2}e^{-2\omega t_*}\right)^{-\frac{1}{2}}\,.
\end{equation}   
It is remarkable that here we have $\delta t \delta x_{||}=\delta t' \delta x'_{||}$ just as in the flat case. The difference is that these effects are decreasing in time vanishing in the limit of $t_* \to \infty$.

These example show how interesting may be the kinematics of the free motion on the de Sitter spacetime. However, here we considered only simple particular examples but we believe that it deserves to investigate this whole complex phenomenology looking for observable effects in our expanding universe.

\subsection{The field of a freely falling electric charge}
\label{sec:6.2}

The next problem which was less studied so far \cite{mos1,mos2} is the electromagnetic field of a massive charged particle freely falling on the expanding portion of the de Siter manifold. A particle of mass $m$  carying the electric charge $q$ produces a Coulomb field in its rest frame $O'$. Then the problem is  how this field is measured by the fixed observer $O$ with respect to which this particle moves with the momentum $\vec{P}$. This example is useful since it has a simple solution in Minkowski spacetime such that we can verify easily the flat limit \cite{LL}.

We start with the  Euclidean chart of $O'$ where the charged particle of mass $m$ stays at rest in ${\vec{x}}' =0$. Here the electromagnetic potential has the same form as in the Minkowski spacetime since the Maxwell equations are invariant under conformal transformations.  In the chart $\{t',{\vec{x}}'\}$ of $O'$ we obtain similar components performing the transformation (\ref{EdS}) and transforming covariantly the electromagnetic potential. Thus we obtain similar formulas
\begin{eqnarray}
&&A'_{(c) 0}(x'_c)=\frac{q}{|\vec{x}'_c|}\,, \quad A'_{(c) i}(x'_c)=0\,\nonumber\\
&\to&  A'_{0}(x')=\frac{q}{|{\vec{x}}'|}\,, \qquad ~~A'_{i}(x')=0\,,\label{Coul}
\end{eqnarray}  
giving the Coulomb field in the mobile frame with de Sitter-Painlev\' e coordinates. Our goal is to calculate this field in the chart $\{t, \vec{x}\}$ of the fixed observer $O$.

The coordinates of the frames $O'$ and $O$ are related through the Lorenzian  isometry $x=\phi_{{\frak g}(\vec{P})}(x')$ given by Eqs. (\ref{Eq1A}) and (\ref{Eq2A}). Then it is obvious that the electromagnetic potential in the frame $O$ has to be calculated according to the general rule 
\begin{eqnarray}
A_{\mu}(x)&=&\frac{\partial x^{\prime\, \nu}}{\partial x^{\mu}}A'_{\nu}(x')\nonumber\\
&=&\frac{\partial \phi_{{\frak g}(-\vec{P})}^{\nu}(x)}{\partial x^{\mu}}A'_{\nu}(\phi_{{\frak g}(-\vec{P})}(x))\,,\label{AAA}
\end{eqnarray} 
where we must use the inverse isometry $x'=\phi_{{\frak g}(-\vec{P})}(x)$. Indeed, from Eqs. (\ref{Coul}) and (\ref{AAA}) we obtain the expression   
\begin{equation}\label{pot}
A_{\mu}(x)=
 \frac{\partial t'(t,\vec{x})}{\partial x^{\mu}}\frac{q}{|{\vec{x}}'(t,{\vec{x}}) |}\,,
\end{equation}
depending only on the functions (\ref{Eq1B}) and (\ref{Eq2B}). We calculate first the quantity 
\begin{eqnarray}
R&=&|{\vec{x}}'(t,{\vec{x}})|\nonumber\\
&=&\left[\frac{1}{m^2}\left(E \Theta -\vec{x}\cdot \vec{P}\right)^2+{\vec{x}}^2-\Theta^2\right]^{\frac{1}{2}},\label{RdS}
\end{eqnarray} 
and then, by using the derivatives of the function (\ref{Eq1B}), we obtain the definitive result
\begin{eqnarray}
A_0&=&\frac{e^{\omega t}E-\omega (E-m) \Theta}{me^{\omega t}-{\omega}{\vec{x}}\cdot \vec{P} +\omega (E-m)\Theta} \,  \frac{q}{R}\,,\\
A_i &=&\frac{-P^i+ \omega x^i (E-m) e^{-\omega t}}{me^{\omega t}-{\omega}{\vec{x}}\cdot \vec{P} +\omega (E-m)\Theta}\,   \frac{q}{R}\,.
\end{eqnarray}
Moreover, we observe that from Eq. (\ref{pot}) we may deduce the form of the field strength, 
\begin{eqnarray}
F_{\mu\nu}&=&A_{\mu,\nu}-A_{\nu,\mu}\nonumber\\
&=&-\frac{1}{R^2}\left(A_{\mu}R\partial_{\nu}R-A_{\nu}R\partial_{\mu}R\right)\,,\label{Fstr}
\end{eqnarray}
where $R$ is given by Eq. (\ref{RdS}) and, consequently, we have:
\begin{eqnarray}
R\partial_t R&=&\frac{1}{m^2}\left({\vec{P}\,}^2\Theta -E \vec{P}\cdot \vec{x}\right)\left(e^{\omega t}-\omega\Theta\right)\\
R\partial_i R&=&\frac{1}{m^2}\left(E\Theta-\vec{P}\cdot\vec{x}\right)\left(Ee^{-\omega t}\omega x^i-P^i\right)\nonumber\\
&+&x^i-\omega x^i\Theta e^{-\omega t}\,.
\end{eqnarray}

Furthermore, we verify that in the limit of $\vec{P} \to 0$, when $E\to m$ and $R\to |\vec{x}|$,  we recover the form of the potential in the rest frame,
\begin{equation}
A_0 \to \frac{q}{|\vec{x}|}\,, \quad A_i = 0\,. 
\end{equation} 
However, this result was expected since it is somewhat trivial. More interesting is to calculate the flat limit when $\omega \to 0$. Then $\Theta \to t$ and 
\begin{equation}
R\to R_0=\left[\frac{1}{m^2}\left(E t -\vec{x}\cdot \vec{P}\right)^2+{\vec{x}}^2-t^2\right]^{\frac{1}{2}}\,,
\end{equation}
such that we recover just the potentials of a charged partcle of momentum $\vec{P}$  moving in Minkowski spacetime \cite{LL},
\begin{equation}\label{flat}
A_0(x)\to \frac{E}{m} \frac {q}{R_0}\,, \quad A_i(x)\to-\frac{P^i}{m} \frac{q}{R_0}\,,
\end{equation}
but written in terms of conserved energy and momentum instead of velocity. Notice that the sign of Eq. (\ref{flat}b) is due to the fact that here we calculated the covariant components. The corresponding contravariant components in Minkowski spacetime give the vector $\vec{A}$ which is oriented along the direction of $\vec{P}$.  

These tests convince us that the potential (\ref{pot}) and the field strength (\ref{Fstr}) derived here are correct being able to lead to new interesting physical results. However, their form is quite complicated such that the study of the specific new effects is difficult requiring algebraic and numeric methods on computer  that exceed the present framework.   

\subsection{Binary fission}
\label{sec:6.3}

We assume now that our particle of mass $m$, staying at rest  in ${\vec{x}}'=0$, explodes at  time $t_c'$ splitting in two fragments $(m_{(+)}, \vec{P}_{(+)}')$ and  $(m_{(-)}, \vec{P}_{(-)}')$ whose momenta with respect to $O'$ are $\vec{P}_{(+)}'=-\vec{P}_{(-)}' =\vec{p}$, complying with the usual conservation rule. Now the problem is to find the corresponding  momenta $\vec{P}_{(+)}$ and $\vec{P}_{(-)}$ that may be measured by the fixed observer $O$ with respect to which the exploding particle had the initial momentum $\vec{P}$. 

The calculation must be done in Euclidean charts where we have already the transformation rule  (\ref{KKg}) among the conserved quantities. Moreover, we assume that the measurement, which is strongly dependent on time, is performed at  initial time $t'_c$ so that the geodesics of the both fragments have the same initial condition, $(t'_c, 0)$, in $O'$. The corresponding initial point in $O$ has the coordinates $(t_{c*}, \vec{x}_{c*})$ which satisfy Eqs. (\ref{Eq1}) and (\ref{Eq2}) for $\vec{x}'_c=0$ that read now
\begin{eqnarray}
t_{c*}&=&t_c(t_c',0)=\frac{2mt'_c}{E+m-(E-m)\omega^2 {t_c'}^2}\,,\label{Eq1F}\\
\vec{x}_{c*}&=&\vec{x}_c(t_c', 0)=\frac{\vec{P}}{m}\frac{1-\omega^2 {t'_c}^2}{E+m-(E-m)\omega^2 {t_c'}^2}\,.\label{Eq2F}
\end{eqnarray}
Notice that, according to Eqs. (\ref{EdS}), $t_{c*}$ defined above and the proper time $t_*$ defined by Eq. (\ref{Eq1C}) are related as in Eq. (\ref{EdS}a)  (i. e. $t_{c*}=-\frac{1}{\omega}\exp(-\omega t_*)$). Thus we fixed the coordinates of the explosion of the particle of mass $m$ in both frames taking into account that these represent the initial conditions of the geodesic trajectories of the resulted fragments as observed by $O$ and $O'$.  

Furthermore, we focus on the first fragment observing that for  ${\vec{x}_c}'=0$ its conserved quantities become  ${\vec{L}}'_{(+)}=0$,  ${\vec{Q}}'_{(+)}=\omega^2 {t'_c}^2 \vec{p}$ and
\begin{equation}
E'_{(+)}=\sqrt{m_{(+)}^2+p^2 \omega^2 {t_c'}^2}
\end{equation} 
such that the matrix (\ref{KK}) takes now the form 
\begin{eqnarray}
&&{\cal K}_{(+)}(t_c',0,\vec{p})\nonumber\\ 
&&~~~~=\left(
\begin{array}{ccccc}
0& -\alpha_- p^1&-\alpha_- p^2&-\alpha_-  p^3&E_{(+)}'\\
\alpha_- p^1&0&0&0&-\alpha_+ p^1\\
\alpha_- p^2&0&0&0&-\alpha_+ p^2\\
\alpha_- p^3&0&0&0&-\alpha_+ p^3\\
-E_{(+)}'&\alpha_+ p^1&\alpha_+ p^2&\alpha_+ p^3&0
\end{array}\right)\,,\label{KK1}
\end{eqnarray} 
where
\begin{equation}
\alpha_{\pm}=\frac{1}{2}(1\pm \omega^2 {t_c'}^2)\,.
\end{equation}
With these ingredients, we intend to calculate the conserved quantities of the both fragments  measured in the fixed frame $O$. This can be done by using the transformation (\ref{KKg}) with the above new initial conditions, 
\begin{equation}\label{KKg1}
{\cal K}_{(\pm)}(t_{c*},\vec{x}_{c*},{\vec{P}_{(\pm)}})=\overline{\frak g}(\vec{P})\,{\cal K}_{(\pm)}(t_c', 0,\pm\vec{p}\,)\,\overline{\frak g}(\vec{P})^T\,.
\end{equation}
This problem is difficult but can be solved resorting to suitable algebraic codes on computer. Thus we derive the momenta observed in $O$ of both fragments as functions of $t'_c$ bearing in mind that this depends on $t_*$ as it results from Eqs.  (\ref{EdS}) and (\ref{ttst}) or solving directly Eq. (\ref{Eq1F}). Performing this substitution, after a few manipulation, we find first that, 
\begin{equation}
E'_{(\pm)}=\left[m_{(\pm)}^2+\frac{p^2 \left(\frac{E}{m}+1\right)^2 e^{-2\omega t_*}}{\left(1+\sqrt{1+\frac{P^2}{m^2}e^{-2\omega t_*}}\right)^2}\right]^{\frac{1}{2}}\,, 
\end{equation}
and then we obtain the final result,
\begin{eqnarray}
\vec{P}_{(\pm)}&=&\frac{E'_{(\pm)}}{m}\vec{P} \pm\frac{1}{2}\left(\frac{E}{m}+1\right)\vec{p}\\
&\pm &\frac{(\frac{E}{m}+1)\left(2(\vec{p}\cdot \vec{P})\vec{P}-P^2\vec{p}\right)e^{-2\omega t_*}}{2m^2\left(1+\sqrt{1+\frac{P^2}{m^2} e^{-2\omega t_*}}\right)^2}
\,.\label{Pfix}
\end{eqnarray}
The corresponding energies measured in $O$ read
\begin{eqnarray}
E_{(\pm)}&=&\frac{E E'_{(\pm)}}{m}\pm\frac{\vec{P}\cdot\vec{p}}{m}\nonumber\\
&\times & \frac{1+\sqrt{1+\frac{P^2}{m^2} e^{-2\omega t_*}}+\frac{E}{m}\left(\frac{E}{m}+1\right)e^{-2\omega t_*}}{\left(1+\sqrt{1+\frac{P^2}{m^2}e^{-2\omega t_*}}\right)^2}\,.\label{Efix}
\end{eqnarray}
Hereby we can verify that in the flat limit, for $\omega \to 0$, we recover the well-known result in Minkowski spacetime presented briefly with our notations in the Appendix B. 

As stated before,  $t_*$ is the time when $O$ observes the explosion of the particle of mass $m$ in $\vec{x}_*$.  Therefore, after this moment,  the trajectories of the resulted fragments observed by $O$ are geodesics with this initial condition and momenta given by Eq. (\ref{Pfix}). For $t\ge t_*$ their equations read
\begin{eqnarray}
&&\vec{x}_{(\pm)}(t)= \vec{x}_* e^{\omega (t- t_*)}+\frac{\vec{P}_{(\pm)}e^{\omega t}}{\omega {\vec{P}_{(\pm)}}^2}\nonumber\\
&&\times\left(\sqrt{m_{(\pm)}^2 + {\vec{P}_{(\pm)}}^2 e^{-2\omega t_*}}- \sqrt{m_{(\pm)}^2 + {\vec{P}_{(\pm)}}^2 e^{-2\omega t}}\right)\,.\nonumber\\
\end{eqnarray}

Finally, we note that the method presented here may be used for analyzing the kinematics of any collision or nuclear reaction on de Sitter spacetime regardless the frames where these are observed.   However, in the weak gravitational field of our expanding universe it is less probable to observe the influence of gravity since the first corrections in Eqs. (\ref{Pfix}) and (\ref{Efix}) are of the order $\omega^2$. 
    
\section{Concluding remarks}
\label{sec:7}

In this paper we completed the Nachtmann  boosting method of introducing coordinates on de Sitter spacetimes  with special gauge transformations giving rise to desired conserved quantities. We obtained thus an effective framework for studying the relative geodesic motion in different local local charts that play here the role of the inertial frames of special relativity.  In this manner  we succeeded to define the natural rest frames of the massive mobiles finding the Lorenzian transformations among these frames and other arbitrary ones.  The applications presented here reveal the possibilities and perspectives of our approach in studying classical relativistic effects on the de Sitter spacetime. 

On the other hand, we expect to obtain  more interesting results in large domains of quantum theory, starting with the  representation theory of  the covariant free fields up to complex processes involving interacting quantum field in gravitational fields or even in investigating how the quantum matter gives rise to gravity.  However, our hope is of finding new observable quantum effects improving thus our knowledge  in astrophysics and cosmology. 

\appendix

\section{General isometry}\label{App:A}

Calculating explicitly the matrix  
\begin{equation}
{\frak g}_*={\frak b}(t_0,\vec{x}_{c0},\vec{P}){\frak b}({t_0'},\vec{x}'_{c0},0)^{-1}
\end{equation}
for arbitrary initial conditions we find the following isometry transformations
\begin{eqnarray}
&&t_c(t'_c,\vec{x}'_c) =\frac{t'_c }{\Delta'}\,,\\
&&\vec{x}_c(t'_c,\vec{x}'_c)=\vec{x}'_{c0} +\frac{1}{\Delta'}\left\{\vec{x}'_c-\vec{x}'_{c0}\right.\nonumber\\
&&\left.~~~~ + \vec{n}_p\left[\frac{E_{t_0}-m}{2m}\, \vec{n}_p\cdot(\vec{x}'_c-\vec{x}'_{c0})\right.\right.\nonumber\\
&&\left.\left.~~~~ +\frac{\omega}{2m}\frac{{t_{c0}}}{{t'_{c0}}}\left({t_{c0}}^2-{t'_c}^2+(\vec{x}'_c-\vec{x}'_{c0})^2\right)\right]\right\}\,,
\end{eqnarray}
where we denote  $E_{{t_{c0}}}=\sqrt{m^2+{P}^2\omega^2 {t_{c0}}^2}$,  $\vec{n}_p=\frac{\vec{P}}{P}$ and
\begin{eqnarray}
\Delta'&=&1+\frac{\omega} {m}\, \vec{P}\cdot(\vec{x}'_c-\vec{x}'_{c0})\nonumber\\
&+& \frac{E_{{t_{c0}}}-m}{2m}\left[\frac{t'_{c0}}{t_{c0}}-\frac{t'^2-(\vec{x}'_c-\vec{x}'_{c0})^2}{{t_{c0}}t_{c0}'}\right]\,.
\end{eqnarray}
For $\vec{P}=0$  we must have ${\frak g}={\frak e}$ but we obtain 
\begin{equation}
t_c' t_{c0}=t_c t_{c0}'\,, \quad
t_{c0}(\vec{x}'_{c}-\vec{x}'_{c0})=t'_{c0}(\vec{x}_c-\vec{x}_{c0})\,,
\end{equation}    
such that  $t_{c0}'=t_{c0}$ and $\vec{x}'_{c0}=\vec{x}_{c0}$ become mandatory conditions.  

\section{Binary fission in flat spacetime}\label{App:B}

The problem of section \ref{sec:6.3} in Minkowski spacetime is solved by using the Lorentz boost $L(\vec{P})$ extracted from Eq. (\ref{Lorbust}) that has the form   
\begin{equation}
{\frak g}(\vec{P})=\left(
\begin{array}{cc}
L(\vec{P})&0\\
0&1
\end{array}\right)\,.
\end{equation} 
Then by applying this boost on the four-momenta components $(E'_{(\pm)}, \pm \vec{p})$ of the two fragments we obtain
\begin{eqnarray}
E_{\pm}&=&\frac{1}{m}\left(EE'_{(\pm)}+\vec{P}\cdot\vec{p}\right)\,,\\
\vec{P}_{(\pm)}&=&\frac{E'_{(\pm)}}{m}\vec{P}\pm \frac{\vec{P}\cdot \vec{p}}{m(E+m)} \vec{P} \pm \vec{p}\,,
\end{eqnarray}
where now $E'_{(\pm)}=\sqrt{m_{(\pm)}^2 + p^2}$.

\end{document}